\documentclass[referee]{raa}            % referee version: for submission
\usepackage{graphicx,times}             %for PS/EPS graphics inclusion, new
\usepackage{natbib}
\usepackage{amssymb,amsmath}
\bibpunct{(}{)}{;}{a}{}{,}

\begin{document}

   \title{New Photometric Investigation of the Low-Mass-Ratio Contact Binary Star V1853 Orionis
%\footnotetext{$*$ Supported by the National Natural Science Foundation of China.}
}
%   \subtitle{I. Place Your Subtitle Here}

   \volnopage{Vol.0 (201x) No.0, 000--000}      %%preserved for Editor. DOn't remove!
   \setcounter{page}{1}           %%starting page, preserved for Editor. DOn't remove!

   \author{Jia-Jia He
      \inst{1,2,3}
   \and Sheng-Bang Qian
      \inst{1,2,3,4}
   \and B. Soonthornthum
      \inst{5}
   \and A. Aungwerojwit
      \inst{6}
   \and Niang-Ping Liu
      \inst{1,2,3}
   \and T. Sarotsakulchai
      \inst{1,2,3,4}
   }

   \institute{Yunnan Observatories, Chinese Academy of Sciences (CAS), P. O. Box 110, 650216 Kunming, China; {\it hjj@ynao.ac.cn}\\
        \and
             Key Laboratory of the Structure and Evolution of Celestial Objects, Chinese Academy of Sciences, P. O. Box 110, 650216 Kunming, China\\
        \and
             Center for Astronomical Mega-Science, Chinese Academy of Sciences, Beijing 100012, China\\
        \and
             University of Chinese Academy of Sciences, Yuquan Road 19\#, Sijingshang Block, 100049 Beijing, China\\
        \and
             National Astronomical Research Institute of Thailand, 191 Siriphanich Bidg. 2nd Fl. Huay Kaew Rd. Suthep District, Muang, Chiang Mai 50200, Thailand\\
        \and
             Department of Physics, Faculty of Science, Naresuan University, Phitsanulok 65000, Thailand\\
        }

   \date{Received~~20xx month day; accepted~~20xx~~month day}

\abstract{Four-color charge-coupled device (CCD) light curves in $B$, $V$, $Rc$ and $Ic$ bands of the total-eclipsing binary system, V1853 Ori, are presented. By comparing our light curves with those published by previous investigators, it is detected that the O'Connell effect on the light curves is disappeared. By analyzing those multi-color light curves with the Wilson-Devinney code (W-D code), it is discovered that V1853 Ori is an A-type intermediate-contact binary with a degree of contact factor of $f=33.3\%(3.7\%)$ and a mass ratio of $q=0.1896(0.0013)$. Combining our 10 new determined times of light minima together with the others published in the literature, the period changes of the system is investigated. We found that the general trend of the observed-calculated $(O-C)$ curve shows a downward parabolic variation that corresponds to a long-term decrease in the orbital period with a rate of $dP/dt=-1.96(0.46)\times{10^{-7}}$ d yr$^{-1}$. The long-term period decrease could be explained by mass transfer from the more-massive component to the less-massive one. By combining our photometric solutions with the Gaia DR 2 data, absolute parameters were derived as $M_{1}$ = 1.20 M$_{\odot}$, $M_{2}$ = 0.23 M$_{\odot}$, $R_{1}$ = 1.36 R$_{\odot}$, and $R_{2}$ = 0.66 R$_{\odot}$. The long-term period decrease and the intermediate-contact configuration suggest that V1853 Ori will evolve into a high fill-out overcontact binary.
}

   \authorrunning{J.-J. He et al. }         %author_head in even pages
   \titlerunning{Photometric Investigation of V1853 Ori}  % title_head in odd pages

   \maketitle
%% The author head (on even pages) and the title head (on odd pages) will be
%% automatically extracted from \author{} and \title{}. Whenever the title is too long,
%% you will be asked to supply a shorter one by inserting either \authorrunning{} or
%% \titlerunning{} before \maketitle. Anyway, you can specify your own heads.
%%
%%
%% Note: In the following text body of your manuscript, please note several differences from
%%       other major journals:
%% (1) \subsection{Please Capitalize the First Letter of Each Notional Word in Subsection Title}
%% (2) Please Capitalize the First Letter of Each Notional Word in all tables' captions

%
%________________________________________________ sections below
%

\section{Introduction}

V1853 Ori (GSC 01283-00053 = NSVS 9553026 = ASAS 051305+155812) was discovered as a variable star by the Robotic Optical Transient Search Experiment (ROTSE)-I telescope by \cite{2006AJ....131..621G}. They pointed out it was a contact binary candidate by using the observed period-color relation and the binary nature was confirmed by visual examination of the light curves. Later, a total of 236 measurements in $V$ and $R$ bands were obtained by \cite{2007IBVS.5799....1.}. They classified V1853 Ori as an EW-type binary and gave the first linear ephemeris,
 \begin{equation}
Min.I(HJD)=2454066.5778+0^{\rm d}.383004\times{E},\label{linear ephemeris1}
 \end{equation}
based on 12 times of minima with the (ROTSE)-I data. Their $V-R$ color curve showed no variation. \cite{2011AJ....142..117S} published $UBVRcIc$ light curves that were obtained on December 26, 29, 30, and 31, 2017 at Lowell Observatory with the Lowell 31 inch reflector. Their light curves showed positive O'Connell effect (the magnitudes at phase 0.25 are lower than magnitudes at phase 0.75, \citealt{1951PRCO....2...85O}) in $B$, $V$ and $Rc$ bands but negative O'Connell effect in $U$ and $Ic$ bands. The asymmetry light curves were explained by two dark spots on the primary star. Their photometric analysis with W-D code showed that this system is an extreme mass ratio W-type overcontact binary ($q=0.20$). The derived fill-out factor $~$$50\%$ revealed that it approaches its final stages of binary star evolution. These properties indicate that V1853 Ori is an interesting target for further investigations. Both components in the binary may merge into a rapid-rotating single star when it meets the more familiar criterion that the orbital angular momentum is less than 3 times the total spin angular momentum, i.e., Jorb $<$ 3Jrot (\citealt{1980A&A....92..167H}). Therefore, it is a progenitor candidate of luminous red novae (e.g., \citealt{2016RAA....16...68Z}).

\cite{2011AJ....142..117S} revised the linear ephemeris as,
\begin{equation}
Min.I(HJD)=2454066.57858+0^{\rm d}.38300155\times{E}.\label{linear ephemeris2}
\end{equation}
They showed that the period of V1853 Ori is not variable. Recently, V1853 Ori was observed by LAMOST (the Large Sky Area Multi-Object Fiber Spectroscopic Telescope) spectroscopic survey on October 15, 2014. Stellar atmospheric parameters of the binary published by \cite{2017RAA....17...87Q} and are listed in Table \ref{lamost}. The distributions of the metallicity ([Fe/H]) and the gravitational acceleration $\log g$ for EWs were given by \cite{2018ApJS..235....5Q}. It is show that the values of V1853 Ori are close to the peaks of the distributions. Moreover the period (0.38 d) of this system is also close to the peak of the period distribution given by \cite{2017RAA....17...87Q}. These reveal that V1853 Ori is a typical EW-type binary. This eclipsing binary was also observed by Gaia DR 2 (\citealt{2018A&A...616A...1G}) and the parallax ($\pi$ = 2.6856(0.0465) mas) was given.

\begin{table}
\caption{LAMOST results of V1853 Ori}\label{lamost}
\begin{center}
\begin{tabular}{llllllll}\hline\hline
Stars       & Obs-Date   & Sp. & $T_{\rm eff}$  & $\log g$   & [Fe/H]\\
            &            &     &  K             & cm/s$^{2}$ &  dex    \\\hline
V1853 Ori   & 2014-10-15 & F5  & 6240           & 4.07       & -0.24 \\
\hline\hline
\end{tabular}
\end{center}
\end{table}

\section{Multi-color CCD Photometric Observations}
\label{sect:Obs}
\begin{figure}
\begin{center}
\includegraphics[width=10cm]{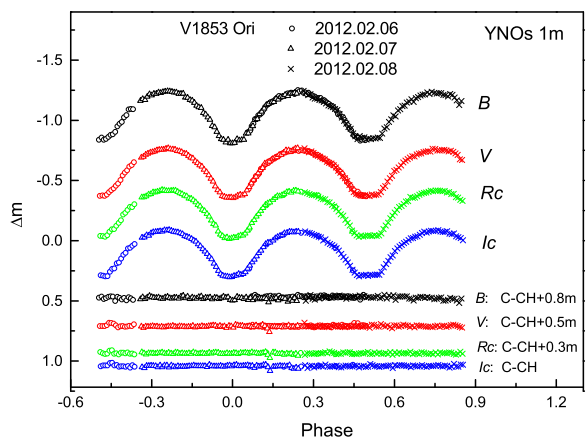}
\caption{CCD photometric light curves in the $B$ and $V$ bands obtained using the 1-m telescope at Yunnan Observatories in 2012. The magnitude difference between the comparison and the check stars are shown in the low pane.}\label{originlc}
\end{center}
\end{figure}

New photometric observations of V1853 Ori were carried out in the $B$, $V$, $Rc$ and $Ic$ bands with the DW436 2048 $\times$ 2048 CCD photometric system attached to the 1-m reflecting telescope at the Yunnan Observatories in China during three nights on February 6, 7 and 8, 2012. The effective field of view of the photometric system was $7' \times 7'$ at the Cassegrain focus. The integration times were 120 s for the $B$ band, 60 s for the $B$ band, 30 s for the $Rc$ and 20 s for the $Ic$ band during our observation, respectively. The comparison star and the check star are in the same field of view of the CCD camera. The coordinates of V1853 Ori, the comparison and the check stars are listed in Table \ref{Coordinates}. PHOT (measure magnitudes for a list of stars) of the aperture photometry package of the IRAF was used to reduce the observed images. The photometric data in $B$, $V$, $Rc$ and $Ic$ bands together with their Heliocentric Julian dates are listed in Table \ref{lcbdata}, Table \ref{lcvdata}, Table \ref{lcrdata} and Table \ref{lcidata}, respectively. The complete CCD light curves in $B$, $V$, $Rc$ and $Ic$ bands with respect to the linear ephemeris,
\begin{equation}
Min.I(HJD)=2455965.11174+0^{\rm d}.38300155\times{E},\label{linear ephemeris}
\end{equation}
are shown in Figure \ref{originlc}. The magnitude difference between the comparison and the check stars are also shown in the low pane of this figure. The epoch in the equation (\ref{linear ephemeris}) is obtained by us and the period is from \cite{2011AJ....142..117S}.

Apart from the 1-m reflecting telescope at the Yunnan Observatories, we also used the 0.4-m telescope at Naresuan University in Thailand and the 60-cm R-C reflect telescope at the Yunnan Observatories in China. By using least-squares parabolic fitting method, 30 CCD times of light minimum averaged into 10 mean eclipse times are listed in Table \ref{Newminimum}.

\begin{figure}
\begin{center}
\includegraphics[width=10cm]{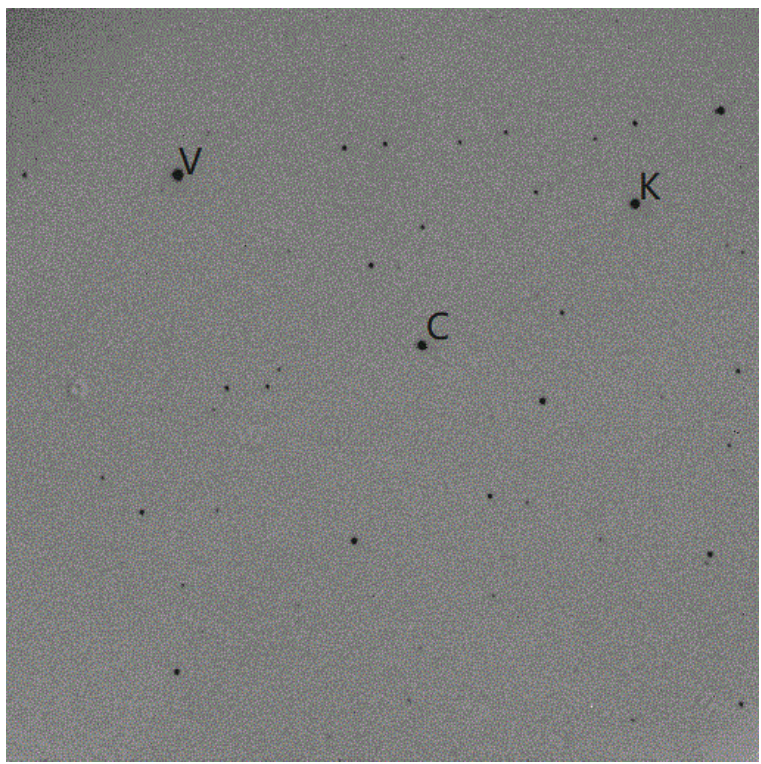}
\caption{One CCD image about V1853 Ori obtained by 1-m reflecting telescope at the Yunnan Observatories including variable (V), comparison (C), and check stars (K).}\label{image}
\end{center}
\end{figure}

\begin{table}
\begin{center}
\caption{Coordinates of V1853 Ori, the comparison, and the check stars}\label{Coordinates}
\begin{tabular}{lll}\hline\hline
Stars           & $\alpha_{2000}$        & $\delta_{2000}$ \\\hline
V1853 Ori       & $05^{\rm h}13^{\rm m}06.069^{\rm s}$ & $+15^\circ58'12.22''$\\
The comparison  & $05^{\rm h}12^{\rm m}48.08^{\rm s}$  & $+15^\circ58'01.8''$\\
The check       & $05^{\rm h}12^{\rm m}49.98^{\rm s}$  & $+15^\circ56'37.4''$\\
\hline\hline
\end{tabular}
\end{center}
\end{table}

\begin{table}
\caption[]{CCD Photometric Data of V1853 Ori in the $B$ band}\label{lcbdata}
\tiny
\begin{center}
\begin{tabular}{lllllllllllllllll}
\hline
JD(Hel.) & $\Delta$(m)&JD(Hel.) & $\Delta$(m)&JD(Hel.) & $\Delta$(m)&JD(Hel.) & $\Delta$(m)&JD(Hel.) & $\Delta$(m)&JD(Hel.) & $\Delta$(m)\\
2459000+ &            &2455900+ &            &2455900+ &           &2455900+ &            &2455900+ &            &2455900+ &             \\
\hline
63.98427 &	-0.901 &	64.07744 &	-1.200 &	64.17892 &	-0.905 &	65.07344 &	-1.048 &	 65.19654 &	-1.220 &	66.09014 &	-0.889 \\
63.98746 &	-0.914 &	64.07981 &	-1.197 &	64.18211 &	-0.949 &	65.07662 &	-1.024 &	 65.19973 &	-1.228 &	66.09333 &	-0.932 \\
63.98982 &	-0.951 &	64.08217 &	-1.183 &	64.18529 &	-0.972 &	65.07981 &	-1.004 &	 65.20292 &	-1.223 &	66.09652 &	-0.943 \\
63.99219 &	-0.973 &	64.08536 &	-1.186 &	64.18848 &	-0.987 &	65.08299 &	-0.954 &	 65.97859 &	-1.244 &	66.09970 &	-0.987 \\
63.99455 &	-1.011 &	64.08773 &	-1.186 &	64.19167 &	-1.025 &	65.08618 &	-0.942 &	 65.98178 &	-1.225 &	66.10289 &	-1.008 \\
63.99692 &	-1.014 &	64.09009 &	-1.162 &	64.19485 &	-1.043 &	65.08937 &	-0.905 &	 65.98496 &	-1.209 &	66.10608 &	-1.042 \\
63.99928 &	-1.041 &	64.09246 &	-1.168 &	64.19804 &	-1.091 &	65.09255 &	-0.869 &	 65.98818 &	-1.216 &	66.10926 &	-1.032 \\
64.00165 &	-1.048 &	64.09482 &	-1.157 &	64.20122 &	-1.094 &	65.09574 &	-0.843 &	 65.99137 &	-1.189 &	66.11245 &	-1.081 \\
64.00402 &	-1.062 &	64.09719 &	-1.143 &	64.20441 &	-1.092 &	65.09893 &	-0.838 &	 65.99456 &	-1.181 &	66.11563 &	-1.093 \\
64.00638 &	-1.086 &	64.09955 &	-1.137 &	64.98103 &	-1.162 &	65.10211 &	-0.817 &	 65.99774 &	-1.180 &	66.11882 &	-1.118 \\
64.00875 &	-1.110 &	64.10192 &	-1.124 &	64.98422 &	-1.177 &	65.10530 &	-0.839 &	 66.00093 &	-1.169 &	66.12201 &	-1.122 \\
64.01193 &	-1.129 &	64.10429 &	-1.118 &	64.98741 &	-1.170 &	65.10849 &	-0.810 &	 66.00411 &	-1.166 &	66.12519 &	-1.148 \\
64.01430 &	-1.138 &	64.10665 &	-1.102 &	64.99059 &	-1.197 &	65.11167 &	-0.810 &	 66.00730 &	-1.152 &	66.12838 &	-1.155 \\
64.01666 &	-1.150 &	64.10984 &	-1.085 &	64.99378 &	-1.216 &	65.11486 &	-0.812 &	 66.01049 &	-1.139 &	66.13157 &	-1.175 \\
64.01903 &	-1.152 &	64.11220 &	-1.053 &	64.99697 &	-1.208 &	65.11804 &	-0.843 &	 66.01367 &	-1.123 &	66.13475 &	-1.152 \\
64.02139 &	-1.161 &	64.11457 &	-1.057 &	65.00015 &	-1.220 &	65.12123 &	-0.833 &	 66.01686 &	-1.117 &	66.13794 &	-1.183 \\
64.02376 &	-1.176 &	64.11693 &	-1.029 &	65.00334 &	-1.220 &	65.12463 &	-0.837 &	 66.02004 &	-1.109 &	66.14113 &	-1.193 \\
64.02612 &	-1.187 &	64.11930 &	-1.026 &	65.00652 &	-1.232 &	65.12781 &	-0.839 &	 66.02323 &	-1.094 &	66.14431 &	-1.205 \\
64.02849 &	-1.202 &	64.12167 &	-1.003 &	65.00971 &	-1.227 &	65.13100 &	-0.890 &	 66.02642 &	-1.085 &	66.14750 &	-1.213 \\
64.03086 &	-1.191 &	64.12403 &	-0.979 &	65.01290 &	-1.235 &	65.13419 &	-0.891 &	 66.02960 &	-1.053 &	66.15083 &	-1.214 \\
64.03322 &	-1.199 &	64.12640 &	-0.959 &	65.01608 &	-1.238 &	65.13737 &	-0.932 &	 66.03279 &	-1.023 &	66.15402 &	-1.232 \\
64.03641 &	-1.202 &	64.12877 &	-0.931 &	65.01927 &	-1.242 &	65.14056 &	-0.952 &	 66.03598 &	-0.995 &	66.15721 &	-1.219 \\
64.03877 &	-1.207 &	64.13114 &	-0.917 &	65.02245 &	-1.240 &	65.14375 &	-1.002 &	 66.03916 &	-0.971 &	66.16039 &	-1.236 \\
64.04114 &	-1.224 &	64.13432 &	-0.877 &	65.02564 &	-1.236 &	65.14693 &	-1.010 &	 66.04235 &	-0.945 &	66.16421 &	-1.227 \\
64.04350 &	-1.222 &	64.13669 &	-0.868 &	65.02883 &	-1.231 &	65.15012 &	-1.053 &	 66.04554 &	-0.930 &	66.16740 &	-1.223 \\
64.04587 &	-1.230 &	64.13905 &	-0.857 &	65.03201 &	-1.232 &	65.15331 &	-1.073 &	 66.04872 &	-0.891 &	66.17058 &	-1.226 \\
64.04824 &	-1.228 &	64.14142 &	-0.854 &	65.03520 &	-1.221 &	65.15649 &	-1.097 &	 66.05191 &	-0.868 &	66.17377 &	-1.226 \\
64.05060 &	-1.219 &	64.14378 &	-0.831 &	65.03839 &	-1.221 &	65.15968 &	-1.127 &	 66.05509 &	-0.843 &	66.17696 &	-1.212 \\
64.05297 &	-1.228 &	64.14615 &	-0.840 &	65.04157 &	-1.222 &	65.16286 &	-1.136 &	 66.05828 &	-0.869 &	66.18014 &	-1.209 \\
64.05533 &	-1.252 &	64.14852 &	-0.852 &	65.04476 &	-1.190 &	65.16718 &	-1.149 &	 66.06147 &	-0.855 &	66.18333 &	-1.198 \\
64.05770 &	-1.240 &	64.15088 &	-0.834 &	65.04794 &	-1.187 &	65.17037 &	-1.152 &	 66.06465 &	-0.859 &	66.18652 &	-1.218 \\
64.06088 &	-1.247 &	64.15607 &	-0.839 &	65.05113 &	-1.170 &	65.17356 &	-1.179 &	 66.06784 &	-0.841 &	66.18970 &	-1.212 \\
64.06325 &	-1.230 &	64.15967 &	-0.856 &	65.05432 &	-1.163 &	65.17674 &	-1.173 &	 66.07103 &	-0.849 &	66.19289 &	-1.189 \\
64.06561 &	-1.232 &	64.16285 &	-0.838 &	65.05750 &	-1.145 &	65.17993 &	-1.193 &	 66.07421 &	-0.843 &	66.19607 &	-1.166 \\
64.06798 &	-1.216 &	64.16604 &	-0.845 &	65.06069 &	-1.122 &	65.18312 &	-1.197 &	 66.07740 &	-0.851 &	66.19926 &	-1.127 \\
64.07035 &	-1.229 &	64.16922 &	-0.863 &	65.06388 &	-1.108 &	65.18630 &	-1.211 &	 66.08058 &	-0.849 &	66.20245 &	-1.160 \\
64.07271 &	-1.215 &	64.17241 &	-0.879 &	65.06706 &	-1.104 &	65.19017 &	-1.202 &	 66.08377 &	-0.856 &		     &         \\
64.07508 &	-1.210 &	64.17560 &	-0.885 &	65.07025 &	-1.061 &	65.19336 &	-1.223 &	 66.08696 &	-0.853 &		     &         \\
\hline
\end{tabular}
\end{center}
\end{table}

\begin{table}
\caption[]{CCD Photometric Data of V1853 Ori in the $V$ band}\label{lcvdata}
\tiny
\begin{center}
\begin{tabular}{lllllllllllllllll}
\hline
JD(Hel.) & $\Delta$(m)&JD(Hel.) & $\Delta$(m)&JD(Hel.) & $\Delta$(m)&JD(Hel.) & $\Delta$(m)&JD(Hel.) & $\Delta$(m)&JD(Hel.) & $\Delta$(m)\\
2455900+ &            &2455900+ &            &2455900+ &           &2455900+ &            &2455900+ &            &2455900+ &             \\
\hline
63.98544 &	-0.429 &	64.07861 &	-0.717 &	64.18009 &	-0.457 &	65.07460 &	-0.569 &	 65.19771 &	-0.753 &	66.09131 &	-0.439  \\
63.98862 &	-0.466 &	64.08098 &	-0.712 &	64.18327 &	-0.475 &	65.07779 &	-0.545 &	 65.20090 &	-0.765 &	66.09450 &	-0.452  \\
63.99099 &	-0.494 &	64.08334 &	-0.714 &	64.18646 &	-0.504 &	65.08097 &	-0.515 &	 65.20408 &	-0.769 &	66.09768 &	-0.491  \\
63.99336 &	-0.504 &	64.08653 &	-0.700 &	64.18965 &	-0.543 &	65.08416 &	-0.491 &	 65.97975 &	-0.768 &	66.10087 &	-0.519  \\
63.99572 &	-0.537 &	64.08889 &	-0.688 &	64.19283 &	-0.567 &	65.08735 &	-0.451 &	 65.98294 &	-0.746 &	66.10406 &	-0.538  \\
63.99809 &	-0.545 &	64.09126 &	-0.684 &	64.19602 &	-0.585 &	65.09053 &	-0.420 &	 65.98613 &	-0.735 &	66.10724 &	-0.579  \\
64.00045 &	-0.559 &	64.09363 &	-0.676 &	64.19921 &	-0.615 &	65.09372 &	-0.402 &	 65.98935 &	-0.741 &	66.11043 &	-0.589  \\
64.00282 &	-0.597 &	64.09599 &	-0.666 &	64.20239 &	-0.629 &	65.09691 &	-0.370 &	 65.99254 &	-0.729 &	66.11362 &	-0.608  \\
64.00519 &	-0.605 &	64.09836 &	-0.654 &	64.20558 &	-0.652 &	65.10009 &	-0.365 &	 65.99572 &	-0.712 &	66.11680 &	-0.624  \\
64.00755 &	-0.626 &	64.10072 &	-0.651 &	64.98220 &	-0.690 &	65.10328 &	-0.357 &	 65.99891 &	-0.705 &	66.11999 &	-0.650  \\
64.00992 &	-0.635 &	64.10309 &	-0.631 &	64.98539 &	-0.697 &	65.10647 &	-0.358 &	 66.00209 &	-0.698 &	66.12318 &	-0.665  \\
64.01310 &	-0.645 &	64.10545 &	-0.633 &	64.98857 &	-0.714 &	65.10965 &	-0.359 &	 66.00528 &	-0.683 &	66.12636 &	-0.677  \\
64.01547 &	-0.662 &	64.10782 &	-0.617 &	64.99176 &	-0.724 &	65.11284 &	-0.354 &	 66.00847 &	-0.676 &	66.12955 &	-0.690  \\
64.01783 &	-0.667 &	64.11100 &	-0.597 &	64.99495 &	-0.733 &	65.11602 &	-0.361 &	 66.01165 &	-0.655 &	66.13273 &	-0.701  \\
64.02020 &	-0.682 &	64.11337 &	-0.583 &	64.99813 &	-0.737 &	65.11921 &	-0.375 &	 66.01484 &	-0.647 &	66.13592 &	-0.700  \\
64.02256 &	-0.684 &	64.11574 &	-0.571 &	65.00132 &	-0.742 &	65.12240 &	-0.381 &	 66.01803 &	-0.641 &	66.13911 &	-0.716  \\
64.02493 &	-0.697 &	64.11810 &	-0.557 &	65.00450 &	-0.751 &	65.12580 &	-0.376 &	 66.02121 &	-0.626 &	66.14229 &	-0.724  \\
64.02729 &	-0.700 &	64.12047 &	-0.531 &	65.00769 &	-0.755 &	65.12898 &	-0.388 &	 66.02440 &	-0.608 &	66.14548 &	-0.726  \\
64.02966 &	-0.712 &	64.12284 &	-0.516 &	65.01088 &	-0.756 &	65.13217 &	-0.419 &	 66.02758 &	-0.580 &	66.14867 &	-0.737  \\
64.03202 &	-0.722 &	64.12520 &	-0.487 &	65.01406 &	-0.756 &	65.13535 &	-0.446 &	 66.03077 &	-0.564 &	66.15200 &	-0.739  \\
64.03439 &	-0.729 &	64.12757 &	-0.473 &	65.01725 &	-0.764 &	65.13854 &	-0.477 &	 66.03396 &	-0.535 &	66.15519 &	-0.743  \\
64.03757 &	-0.732 &	64.12994 &	-0.457 &	65.02044 &	-0.769 &	65.14173 &	-0.506 &	 66.03714 &	-0.518 &	66.15838 &	-0.740  \\
64.03994 &	-0.740 &	64.13230 &	-0.433 &	65.02362 &	-0.760 &	65.14491 &	-0.532 &	 66.04033 &	-0.495 &	66.16156 &	-0.763  \\
64.04231 &	-0.742 &	64.13549 &	-0.419 &	65.02681 &	-0.758 &	65.14810 &	-0.561 &	 66.04352 &	-0.462 &	66.16538 &	-0.762  \\
64.04467 &	-0.744 &	64.13786 &	-0.392 &	65.03000 &	-0.757 &	65.15129 &	-0.586 &	 66.04670 &	-0.430 &	66.16856 &	-0.748  \\
64.04704 &	-0.744 &	64.14022 &	-0.408 &	65.03318 &	-0.746 &	65.15447 &	-0.605 &	 66.04989 &	-0.404 &	66.17175 &	-0.748  \\
64.04940 &	-0.755 &	64.14259 &	-0.375 &	65.03637 &	-0.744 &	65.15766 &	-0.641 &	 66.05308 &	-0.389 &	66.17494 &	-0.753  \\
64.05177 &	-0.751 &	64.14495 &	-0.379 &	65.03955 &	-0.736 &	65.16084 &	-0.650 &	 66.05626 &	-0.384 &	66.17812 &	-0.755  \\
64.05414 &	-0.752 &	64.14732 &	-0.370 &	65.04274 &	-0.724 &	65.16403 &	-0.658 &	 66.05945 &	-0.378 &	66.18131 &	-0.749  \\
64.05650 &	-0.745 &	64.14968 &	-0.376 &	65.04593 &	-0.721 &	65.16835 &	-0.677 &	 66.06263 &	-0.376 &	66.18449 &	-0.741  \\
64.05887 &	-0.752 &	64.15205 &	-0.368 &	65.04911 &	-0.705 &	65.17154 &	-0.687 &	 66.06582 &	-0.378 &	66.18768 &	-0.747  \\
64.06205 &	-0.750 &	64.15724 &	-0.372 &	65.05230 &	-0.691 &	65.17472 &	-0.694 &	 66.06901 &	-0.375 &	66.19087 &	-0.738  \\
64.06442 &	-0.748 &	64.16083 &	-0.376 &	65.05549 &	-0.683 &	65.17791 &	-0.713 &	 66.07219 &	-0.370 &	66.19406 &	-0.725  \\
64.06678 &	-0.748 &	64.16402 &	-0.373 &	65.05867 &	-0.660 &	65.18110 &	-0.715 &	 66.07538 &	-0.383 &	66.19724 &	-0.712  \\
64.06915 &	-0.741 &	64.16720 &	-0.383 &	65.06186 &	-0.645 &	65.18428 &	-0.730 &	 66.07856 &	-0.378 &	66.20043 &	-0.675  \\
64.07151 &	-0.741 &	64.17039 &	-0.403 &	65.06504 &	-0.627 &	65.18747 &	-0.750 &	 66.08175 &	-0.382 &	66.20361 &	-0.671  \\
64.07388 &	-0.734 &	64.17358 &	-0.420 &	65.06823 &	-0.616 &	65.19134 &	-0.751 &	 66.08494 &	-0.385 &		     &          \\
64.07624 &	-0.730 &	64.17676 &	-0.435 &	65.07142 &	-0.590 &	65.19452 &	-0.742 &	 66.08812 &	-0.404 &		     &          \\
\hline
\end{tabular}
\end{center}
\end{table}

\begin{table}
\caption[]{CCD Photometric Data of V1853 Ori in the $Rc$ band}\label{lcrdata}
\tiny
\begin{center}
\begin{tabular}{lllllllllllllllll}
\hline
JD(Hel.) & $\Delta$(m)&JD(Hel.) & $\Delta$(m)&JD(Hel.) & $\Delta$(m)&JD(Hel.) & $\Delta$(m)&JD(Hel.) & $\Delta$(m)&JD(Hel.) & $\Delta$(m)\\
2455900+ &            &2455900+ &            &2455900+ &           &2455900+ &            &2455900+ &            &2455900+ &             \\
\hline
63.98608 &	-0.095 &	64.99878 &	-0.402 &	65.08800 &	-0.126 &	65.17856 &	-0.369 &	 66.04098 &	-0.147 &	66.13020 &	-0.363 \\
64.01056 &	-0.294 &	65.00197 &	-0.401 &	65.09118 &	-0.077 &	65.18175 &	-0.378 &	 66.04416 &	-0.125 &	66.13338 &	-0.356 \\
64.03503 &	-0.380 &	65.00515 &	-0.407 &	65.09437 &	-0.051 &	65.18493 &	-0.388 &	 66.04735 &	-0.084 &	66.13657 &	-0.358 \\
64.05951 &	-0.407 &	65.00834 &	-0.412 &	65.09755 &	-0.038 &	65.18867 &	-0.404 &	 66.05054 &	-0.061 &	66.13975 &	-0.369 \\
64.08399 &	-0.376 &	65.01153 &	-0.421 &	65.10074 &	-0.022 &	65.19199 &	-0.408 &	 66.05372 &	-0.043 &	66.14294 &	-0.381 \\
64.10846 &	-0.281 &	65.01471 &	-0.419 &	65.10393 &	-0.028 &	65.19517 &	-0.400 &	 66.05691 &	-0.033 &	66.14613 &	-0.391 \\
64.13295 &	-0.076 &	65.01790 &	-0.413 &	65.10711 &	-0.015 &	65.19836 &	-0.415 &	 66.06010 &	-0.042 &	66.14931 &	-0.397 \\
64.15788 &	-0.040 &	65.02108 &	-0.413 &	65.11030 &	-0.019 &	65.20154 &	-0.416 &	 66.06328 &	-0.034 &	66.15265 &	-0.399 \\
64.16148 &	-0.039 &	65.02427 &	-0.414 &	65.11349 &	-0.026 &	65.20473 &	-0.409 &	 66.06647 &	-0.040 &	66.15584 &	-0.406 \\
64.16467 &	-0.033 &	65.02746 &	-0.416 &	65.11667 &	-0.024 &	65.98041 &	-0.401 &	 66.06965 &	-0.043 &	66.15902 &	-0.415 \\
64.16785 &	-0.045 &	65.03064 &	-0.417 &	65.11986 &	-0.037 &	65.98359 &	-0.395 &	 66.07284 &	-0.041 &	66.16221 &	-0.411 \\
64.17104 &	-0.074 &	65.03383 &	-0.406 &	65.12305 &	-0.044 &	65.98682 &	-0.387 &	 66.07603 &	-0.039 &	66.16603 &	-0.416 \\
64.17423 &	-0.094 &	65.03701 &	-0.395 &	65.12644 &	-0.043 &	65.99000 &	-0.380 &	 66.07921 &	-0.045 &	66.16921 &	-0.416 \\
64.17741 &	-0.096 &	65.04020 &	-0.387 &	65.12963 &	-0.060 &	65.99319 &	-0.380 &	 66.08240 &	-0.040 &	66.17240 &	-0.415 \\
64.18073 &	-0.124 &	65.04339 &	-0.374 &	65.13282 &	-0.084 &	65.99637 &	-0.376 &	 66.08558 &	-0.043 &	66.17558 &	-0.411 \\
64.18392 &	-0.148 &	65.04657 &	-0.368 &	65.13600 &	-0.110 &	65.99956 &	-0.365 &	 66.08877 &	-0.060 &	66.17877 &	-0.413 \\
64.18711 &	-0.169 &	65.04976 &	-0.350 &	65.13919 &	-0.147 &	66.00274 &	-0.352 &	 66.09196 &	-0.096 &	66.18196 &	-0.400 \\
64.19029 &	-0.218 &	65.05295 &	-0.349 &	65.14238 &	-0.171 &	66.00593 &	-0.345 &	 66.09515 &	-0.125 &	66.18514 &	-0.404 \\
64.19348 &	-0.231 &	65.05613 &	-0.336 &	65.14556 &	-0.201 &	66.00911 &	-0.344 &	 66.09833 &	-0.149 &	66.18833 &	-0.388 \\
64.19667 &	-0.255 &	65.05932 &	-0.320 &	65.14875 &	-0.221 &	66.01230 &	-0.322 &	 66.10152 &	-0.187 &	66.19152 &	-0.389 \\
64.19985 &	-0.281 &	65.06251 &	-0.297 &	65.15193 &	-0.249 &	66.01549 &	-0.317 &	 66.10470 &	-0.208 &	66.19470 &	-0.362 \\
64.20304 &	-0.279 &	65.06569 &	-0.284 &	65.15512 &	-0.272 &	66.01867 &	-0.297 &	 66.10789 &	-0.233 &	66.19789 &	-0.357 \\
64.20623 &	-0.303 &	65.06888 &	-0.258 &	65.15831 &	-0.297 &	66.02186 &	-0.284 &	 66.11108 &	-0.255 &	66.20108 &	-0.340 \\
64.98285 &	-0.360 &	65.07206 &	-0.247 &	65.16149 &	-0.311 &	66.02505 &	-0.265 &	 66.11426 &	-0.279 &	66.20426 &	-0.330 \\
64.98604 &	-0.362 &	65.07525 &	-0.223 &	65.16468 &	-0.318 &	66.02823 &	-0.242 &	 66.11745 &	-0.289 &		     &         \\
64.98922 &	-0.369 &	65.07844 &	-0.201 &	65.16900 &	-0.339 &	66.03142 &	-0.219 &	 66.12064 &	-0.305 &		     &         \\
64.99241 &	-0.384 &	65.08162 &	-0.167 &	65.17219 &	-0.356 &	66.03460 &	-0.208 &	 66.12382 &	-0.325 &		     &         \\
64.99559 &	-0.387 &	65.08481 &	-0.139 &	65.17537 &	-0.370 &	66.03779 &	-0.174 &	 66.12701 &	-0.328 &	         &         \\
\hline
\end{tabular}
\end{center}
\end{table}

\begin{table}
\caption[]{CCD Photometric Data of V1853 Ori in the $Ic$ band}\label{lcidata}
\tiny
\begin{center}
\begin{tabular}{lllllllllllllllll}
\hline
JD(Hel.) & $\Delta$(m)&JD(Hel.) & $\Delta$(m)&JD(Hel.) & $\Delta$(m)&JD(Hel.) & $\Delta$(m)&JD(Hel.) & $\Delta$(m)&JD(Hel.) & $\Delta$(m)\\
2455900+ &            &2455900+ &            &2455900+ &           &2455900+ &            &2455900+ &            &2455900+ &             \\
\hline
63.98650 &	0.233 	& 64.99920 & 	-0.071 	& 65.08841 & 	0.217 	& 65.17898 & 	-0.040 	& 66.04139 & 	0.182 	& 66.13061 & 	-0.020 \\
64.01098 &	0.032 	& 65.00238 & 	-0.072 	& 65.09160 & 	0.250 	& 65.18216 & 	-0.045 	& 66.04458 & 	0.209 	& 66.13380 & 	-0.021 \\
64.03545 &	-0.051 	& 65.00557 & 	-0.074 	& 65.09479 & 	0.278 	& 65.18535 & 	-0.062 	& 66.04777 & 	0.241 	& 66.13698 & 	-0.033 \\
64.05993 &	-0.066 	& 65.00875 & 	-0.079 	& 65.09797 & 	0.291 	& 65.18909 & 	-0.066 	& 66.05095 & 	0.265 	& 66.14017 & 	-0.037 \\
64.08440 &	-0.038 	& 65.01194 & 	-0.077 	& 65.10116 & 	0.296 	& 65.19240 & 	-0.075 	& 66.05414 & 	0.286 	& 66.14336 & 	-0.049 \\
64.10888 &	0.061 	& 65.01513 & 	-0.087 	& 65.10434 & 	0.298 	& 65.19559 & 	-0.073 	& 66.05733 & 	0.296 	& 66.14654 & 	-0.063 \\
64.13337 &	0.246 	& 65.01831 & 	-0.086 	& 65.10753 & 	0.304 	& 65.19877 & 	-0.070 	& 66.06051 & 	0.288 	& 66.14973 & 	-0.074 \\
64.15830 &	0.288 	& 65.02150 & 	-0.089 	& 65.11072 & 	0.301 	& 65.20196 & 	-0.075 	& 66.06370 & 	0.288 	& 66.15307 & 	-0.067 \\
64.16190 &	0.290 	& 65.02469 & 	-0.082 	& 65.11390 & 	0.292 	& 65.20515 & 	-0.074 	& 66.06688 & 	0.284 	& 66.15625 & 	-0.077 \\
64.16508 &	0.296 	& 65.02787 & 	-0.075 	& 65.11709 & 	0.289 	& 65.98082 & 	-0.076 	& 66.07007 & 	0.292 	& 66.15944 & 	-0.082 \\
64.16827 &	0.295 	& 65.03106 & 	-0.071 	& 65.12028 & 	0.277 	& 65.98401 & 	-0.066 	& 66.07326 & 	0.284 	& 66.16263 & 	-0.080 \\
64.17146 &	0.258 	& 65.03424 & 	-0.070 	& 65.12346 & 	0.283 	& 65.98723 & 	-0.063 	& 66.07644 & 	0.285 	& 66.16644 & 	-0.082 \\
64.17464 &	0.234 	& 65.03743 & 	-0.061 	& 65.12686 & 	0.276 	& 65.99042 & 	-0.056 	& 66.07963 & 	0.283 	& 66.16963 & 	-0.087 \\
64.17783 &	0.231 	& 65.04062 & 	-0.054 	& 65.13005 & 	0.262 	& 65.99360 & 	-0.048 	& 66.08281 & 	0.282 	& 66.17281 & 	-0.076 \\
64.18115 &	0.189 	& 65.04380 & 	-0.038 	& 65.13323 & 	0.232 	& 65.99679 & 	-0.042 	& 66.08600 & 	0.284 	& 66.17600 & 	-0.073 \\
64.18434 &	0.182 	& 65.04699 & 	-0.032 	& 65.13642 & 	0.207 	& 65.99997 & 	-0.032 	& 66.08919 & 	0.260 	& 66.17919 & 	-0.072 \\
64.18752 &	0.149 	& 65.05018 & 	-0.025 	& 65.13961 & 	0.179 	& 66.00316 & 	-0.027 	& 66.09237 & 	0.222 	& 66.18237 & 	-0.078 \\
64.19071 &	0.117 	& 65.05336 & 	-0.014 	& 65.14279 & 	0.158 	& 66.00634 & 	-0.001 	& 66.09556 & 	0.193 	& 66.18556 & 	-0.059 \\
64.19390 &	0.100 	& 65.05655 & 	-0.001 	& 65.14598 & 	0.130 	& 66.00953 & 	-0.011 	& 66.09875 & 	0.169 	& 66.18875 & 	-0.055 \\
64.19708 &	0.065 	& 65.05974 & 	0.008 	& 65.14916 & 	0.102 	& 66.01272 & 	0.016 	& 66.10194 & 	0.137 	& 66.19193 & 	-0.060 \\
64.20027 &	0.059 	& 65.06292 & 	0.029 	& 65.15235 & 	0.083 	& 66.01590 & 	0.028 	& 66.10512 & 	0.123 	& 66.19512 & 	-0.037 \\
64.20346 &	0.042 	& 65.06611 & 	0.046 	& 65.15554 & 	0.053 	& 66.01909 & 	0.037 	& 66.10831 & 	0.087 	& 66.19831 & 	-0.029 \\
64.20664 &	0.028 	& 65.06929 & 	0.070 	& 65.15872 & 	0.027 	& 66.02228 & 	0.048 	& 66.11150 & 	0.065 	& 66.20149 & 	-0.022 \\
64.98327 &	-0.026 	& 65.07248 & 	0.090 	& 65.16191 & 	0.016 	& 66.02546 & 	0.063 	& 66.11468 & 	0.047 	& 66.20468 & 	-0.003 \\
64.98645 &	-0.027 	& 65.07567 & 	0.109 	& 65.16509 & 	0.015 	& 66.02865 & 	0.083 	& 66.11787 & 	0.035 	& 	       &           \\
64.98964 &	-0.045 	& 65.07885 & 	0.133 	& 65.16942 & 	-0.001 	& 66.03184 & 	0.110 	& 66.12105 & 	0.021 	& 	       &           \\
64.99282 &	-0.042 	& 65.08204 & 	0.154 	& 65.17260 & 	-0.020 	& 66.03502 & 	0.136 	& 66.12424 & 	0.010 	& 	       &           \\
64.99601 &	-0.062 	& 65.08522 & 	0.191 	& 65.17579 & 	-0.030 	& 66.03821 & 	0.157 	& 66.12743 & 	-0.006  &          &           \\
\hline
\end{tabular}
\end{center}
\end{table}

\begin{table}
\begin{center}
\caption{New CCD times of light minimum for V1853 Ori}\label{Newminimum}
\begin{tabular}{lllllll}\hline
 No.   & JD (Hel.)    & Error (days) &  Method & Min. & Filter  & tel \\\hline
 1     &        2455886.78873 & 0.00049  & CCD & II    & $B$    & Thai-0.4m\\
       &        2455886.78886 & 0.00026  & CCD & II    & $V$    & Thai-0.4m\\
       &        2455886.78917 & 0.00049  & CCD & II    & $Rc$   & Thai-0.4m\\
       &        2455886.78932 & 0.00037  & CCD & II    & $Ic$   & Thai-0.4m\\
 mean  &        2455886.78902 &          & CCD & II    &        & Thai-0.4m\\
 2     &        2455887.74322 & 0.00028  & CCD & I     & $B$    & Thai-0.4m\\
       &        2455887.74400 & 0.00033  & CCD & I     & $V$    & Thai-0.4m\\
       &        2455887.74434 & 0.00018  & CCD & I     & $Rc$   & Thai-0.4m\\
       &        2455887.74448 & 0.00026  & CCD & I     & $Ic$   & Thai-0.4m\\
 mean  &        2455887.74401 &          & CCD & I     &        & Thai-0.4m\\
 3     &        2455964.15425 & 0.00026  & CCD & II    & $B$    & YNOs-1m\\
       &        2455964.15473 & 0.00027  & CCD & II    & $V$    & YNOs-1m\\
 mean  &        2455964.15449 &          & CCD & II    &        & YNOs-1m\\
 4     &        2455965.11158 & 0.00029  & CCD & I     & $B$    & YNOs-1m\\
       &        2455965.11163 & 0.00023  & CCD & I     & $V$    & YNOs-1m\\
       &        2455965.11191 & 0.00022  & CCD & I     & $Rc$   & YNOs-1m\\
       &        2455965.11185 & 0.00028  & CCD & I     & $Ic$   & YNOs-1m\\
 mean  &        2455965.11174 &          & CCD & I     &        & YNOs-1m\\
 5     &        2455966.06929 & 0.00035  & CCD & II    & $B$    & YNOs-1m\\
       &        2455966.06934 & 0.00039  & CCD & II    & $V$    & YNOs-1m\\
       &        2455966.06917 & 0.00032  & CCD & II    & $Rc$   & YNOs-1m\\
       &        2455966.06916 & 0.00031  & CCD & II    & $Ic$   & YNOs-1m\\
 mean  &        2455966.06924 &          & CCD & II    &        & YNOs-1m\\
 6     &        2456603.19052 & 0.00027  & CCD & I     & $I$    & YNOs-1m\\
       &        2456603.18858 & 0.00031  & CCD & I     & $N$    & YNOs-1m\\
       &        2456603.18920 & 0.00020  & CCD & I     & $Rc$   & YNOs-1m\\
mean   &        2456603.18943 &          & CCD &       &        & YNOs-60cm\\
 7     &        2456976.23016 & 0.00035  & CCD & I     & $Ic$   & YNOs-60cm\\
       &        2456976.23196 & 0.00041  & CCD & I     & $Rc$   & YNOs-60cm\\
mean   &        2456976.23106 &          & CCD &       &        & YNOs-60cm\\
 8     &        2457394.08111 & 0.00140  & CCD & I     & $Ic$   & YNOs-60cm\\
       &        2457394.08361 & 0.00043  & CCD & I     & $Rc$   & YNOs-60cm\\
mean   &        2457394.08236 &          & CCD &       &        & YNOs-60cm\\
 9     &        2458100.14243 & 0.00027  & CCD & II    & $Rc$   & YNOs-60cm\\
       &        2458100.14083 & 0.00029  & CCD & II    & $V$    & YNOs-60cm\\
mean   &        2458100.14163 &          & CCD &       &        & YNOs-60cm\\
 10    &        2458109.14383 & 0.00015  & CCD & I     & $V$    & YNOs-1m\\
       &        2458109.14383 & 0.00019  & CCD & I     & $Rc$   & YNOs-1m\\
       &        2458109.14379 & 0.00016  & CCD & I     & $Ic$   & YNOs-1m\\
mean   &        2458109.14382 &          & CCD &       &        & YNOs-1m\\
\hline
\end{tabular}
\end{center}
Tel.: Thai-0.4-m: 0.4 m telescope, Naresuan University, Thailand\\
$~~~~~~~~~$YNOs-1m: 1.0 m R-C reflect telescope, Yunnan Observatories, Chinese Academy of Sciences, China\\
$~~~~~~~~~$YNOs-60cm: 60 cm R-C reflect telescope, Yunnan Observatories, Chinese Academy of Sciences, China\\
\end{table}

\section{Orbital Period Investigation}
\begin{figure}
\begin{center}
\includegraphics[width=10cm]{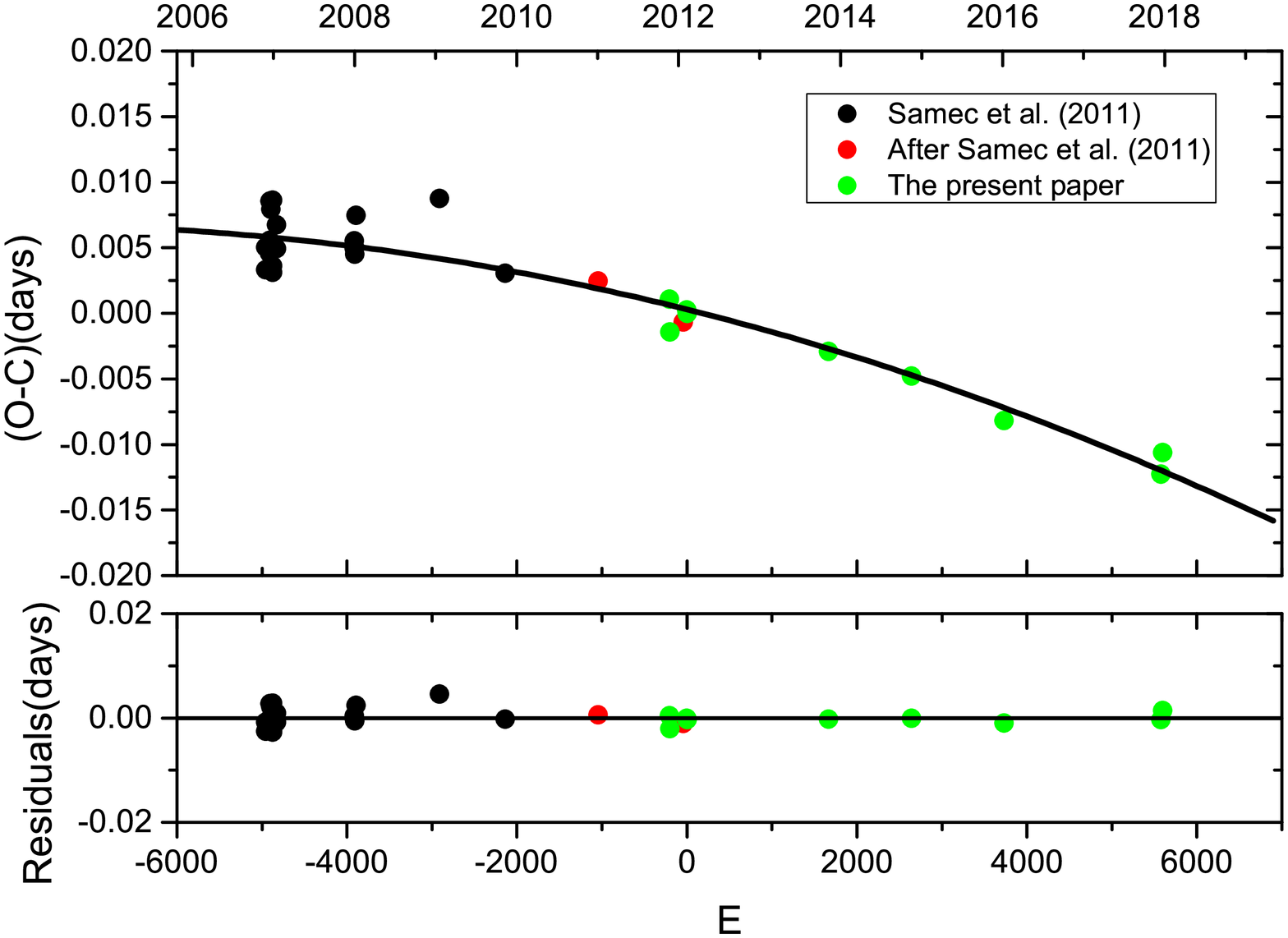}
\caption{$(O-C)$ diagram of the short-period close binary V1853 Ori based on all the primary and secondary available photoelectric and CCD times of light minimum. The quadratic fit ($black$ $solid$ $line$) showed a long-term period decrease. Residuals with respect to the quadratic ephemerides are shown in the $lower$ $panel$. The $black$ $solid$ $dots$ refer to the data from \cite{2011AJ....142..117S}, the $red$ $solid$ $dots$ refer to the data obtained by \cite{2011IBVS.5992....1D} and \cite{2012IBVS.6029....1D} and $green$ $solid$ $dots$ refer to the data obtained by us.}\label{ocimage}
\end{center}
\end{figure}

Since the variable was discovered by \cite{2006AJ....131..621G}, the period changes were study only by \cite{2011AJ....142..117S}, and their study shows no variation. \cite{2011IBVS.5992....1D} and \cite{2012IBVS.6029....1D} published two times of light minimum of V1853 Ori. To study the variations of the orbital period, we collect all the available photoelectric and CCD times of light minimum from the websites and literature. All the primary and the secondary times of light minimum are listed in the Table \ref{O-C}. The $(O-C)$ values (observational times of light minimum-calculational times of light minimum) calculated by equation (\ref{linear ephemeris}) are listed in the third column of Table \ref{O-C} and plotted in the upper panel of Figure \ref{ocimage}. In this figure, the black solid dots refer to the data from \cite{2011AJ....142..117S}, the red solid dots refer to the data obtained by \cite{2011IBVS.5992....1D} and \cite{2012IBVS.6029....1D} and green solid dots refer to the data obtained by us. A least-squares solution yields the following ephemeris,
\begin{equation}
\begin{array}{llll}
(O-C)=2455965.11204(0.00046)+0^{d}.38299992(0^{\rm d}.00000008)\times{E}
         \\-1.03(0.24)\times{10^{-10}}{E^{2}}.\label{epoch}
\end{array}
\end{equation}
With the quadratic term of this ephemeris, a continuous period decrease, at a rate of $dP/dt=-1.96(0.46)\times10^{-7}$ d yr$^{-1}$ is determined. The residuals from equation (\ref{epoch}) are showed in the lower panel of Figure \ref{ocimage} and listed in the column 4 of Table \ref{O-C}.

\begin{table*}
\caption{$(O-C)$ values of light minimum times for V1853 Ori}\label{O-C}
\begin{center}
\begin{tabular}{lclrrrlll}\hline\hline
JD(Hel.)        & Cycles      &$(O-C)$   &Residual      &Reference\\\hline
2454848.6710    & -2915      &  0.00878   &  0.00461    & \cite{2009IBVS.5894....1D}  \\
2455144.9170    & -2141.5    &  0.00308   &  -0.00022   & \cite{2010IBVS.5920....1D}  \\
2454066.3849    & -4957.5    &  0.00334   &  -0.00249   & \cite{2007IBVS.5781....1D}  \\
2454066.5781    & -4957      &  0.00504   &  -0.00079   & \cite{2007IBVS.5781....1D}  \\
2454083.4307    & -4913      &  0.00558   &  -0.00023   & \cite{2007IBVS.5799....1.}  \\
2454083.6212    & -4912.5    &  0.00457   &  -0.00123   & \cite{2007IBVS.5799....1.} \\
2454085.3487    & -4908      &  0.00857   &  0.00276    & \cite{2007IBVS.5799....1.} \\
2454085.5363    & -4907.5    &  0.00467   &  -0.00114   & \cite{2007IBVS.5799....1.} \\
2454090.3271    & -4895      &  0.00795   &  0.00214    & \cite{2007IBVS.5799....1.} \\
2454090.5155    & -4894.5    &  0.00485   &  -0.00095   & \cite{2007IBVS.5799....1.} \\
2454097.2218    & -4877      &  0.00862   &  0.00283    & \cite{2007IBVS.5799....1.} \\
2454097.4078    & -4876.5    &  0.00312   &  -0.00266   & \cite{2007IBVS.5799....1.} \\
2454097.5998    & -4876      &  0.00362   &  -0.00216   & \cite{2007IBVS.5799....1.} \\
2454114.2635    & -4832.5    &  0.00675   &  0.00099    & \cite{2007IBVS.5799....1.} \\
2454114.4532    & -4832      &  0.00495   &  -0.00081   & \cite{2007IBVS.5799....1.} \\
2454464.8998    & -3917      &  0.00513   &  0.00003    & \cite{2011AJ....142..117S}  \\
2454465.6656    & -3915      &  0.00493   &  -0.00016   & \cite{2011AJ....142..117S}  \\
2454465.8577    & -3914.5    &  0.00553   &  0.00043    & \cite{2011AJ....142..117S}  \\
2454466.8142    & -3912      &  0.00452   &  -0.00056   & \cite{2011AJ....142..117S}  \\
2454474.2857    & -3892.5    &  0.00749   &  0.00242    & \cite{2007IBVS.5799....1.}  \\
2455564.68610   & -1045.5    &  0.00248   &  0.00059    & \cite{2011IBVS.5992....1D}  \\
2455947.68450   & -45.5      &  -0.00067  &  -0.00103   & \cite{2012IBVS.6029....1D}  \\
2455886.78902   & -204.5     &  0.00110   &  0.00047    & The present paper  \\
2455887.74401   & -202       &  -0.00142  &  -0.00203   & The present paper  \\
2455964.15449   & -2.5       &  0.00025   &  -0.00004   & The present paper  \\
2455965.11174   & 0          &  0.00000   &  -0.00029   & The present paper  \\
2455966.06924   & 2.5        &  0.00000   &  -0.00028   & The present paper  \\
2456603.18943   & 1666       &  -0.00289  &  -0.00018   & The present paper  \\
2456976.23106   & 2640       &  -0.00477  &  -0.00004   & The present paper  \\
2457394.08236   & 3731       &  -0.00816  &  -0.00094   & The present paper  \\
2458100.14163   & 5574.5     &  -0.01225  &  -0.00026   & The present paper  \\
2458109.14382   & 5598       &  -0.01060  &  0.00144    & The present paper  \\
\hline
\end{tabular}
\end{center}
\end{table*}

\section{Photometric Solutions}

 To understand its geometrical structure and evolutionary state, the $B$, $V$, $Rc$ and $Ic$ light curves shown in the Figure \ref{originlc} were analyzed by using the W-D code (\citealt{1971ApJ....166..605W, 2003Wilson, Wilson2010, 1993AJ....106..2096V}). The light curves published by \cite{2011AJ....142..117S} were asymmetric and showed O'Connell effect that could be explained by dark spot activity on the cool component (e.g., \citealt{2017ApJ...848..131Q}). However, the light curves displayed in Figure \ref{originlc} are almost symmetric and show typical EW type variations, which enables to determine reliable photometric parameters. The magnitudes in $J$, $H$ and $K$ bands (10.46, 10.20, and 10.10\,mag) were obtained from 2MASS (\citealt{2003yCat.2246....0C}). The color indexes $J-H$ = 0.26 and $H-K$ = 0.10\,mag indicate that the spectral type of V1853 Ori is F6 or K3. The corresponding effective temperatures are estimated as 6400\,K or 4800\,K (\citealt{2000asqu.book..381D}), which are quite different. The effective temperature (6240\,K) obtained by LAMOST is in that rang. Therefore, during the solution process, the effective temperature of star 1 was chosen as $T_1=6240$ K. As shown in Figure \ref{originlc}, the depths of both minima are nearly the same indicating the nearly same temperature of both components. Therefore, we take the same values of the gravity-darkening coefficients and the bolometric albedo for both components, i.e., g$_{1}$ = g$_{2}$ = 0.32 (\citealt{1967ZA.....65...89L}) and A$_{1}$ = A$_{2}$ = 0.5 (\citealt{1969AcA....19..245R}). The limb-darkening coefficients were used according to \cite{1990A&A...230..412C} (x and y are the bolometric and bandpass limb-darkening coefficients.). The adjustable parameters include: the orbital inclination ($i$); the mean temperature of star 2 ($T_{2}$); the monochromatic luminosity of star 1 ($L_{1B}$, $L_{1V}$, $L_{1Rc}$, $L_{1Ic}$); and the dimensionless potential of star 1 ($\Omega_{1}=\Omega_{2}$, mode 3 for overcontact configuration).

\begin{figure}
\begin{center}
\includegraphics[width=10cm]{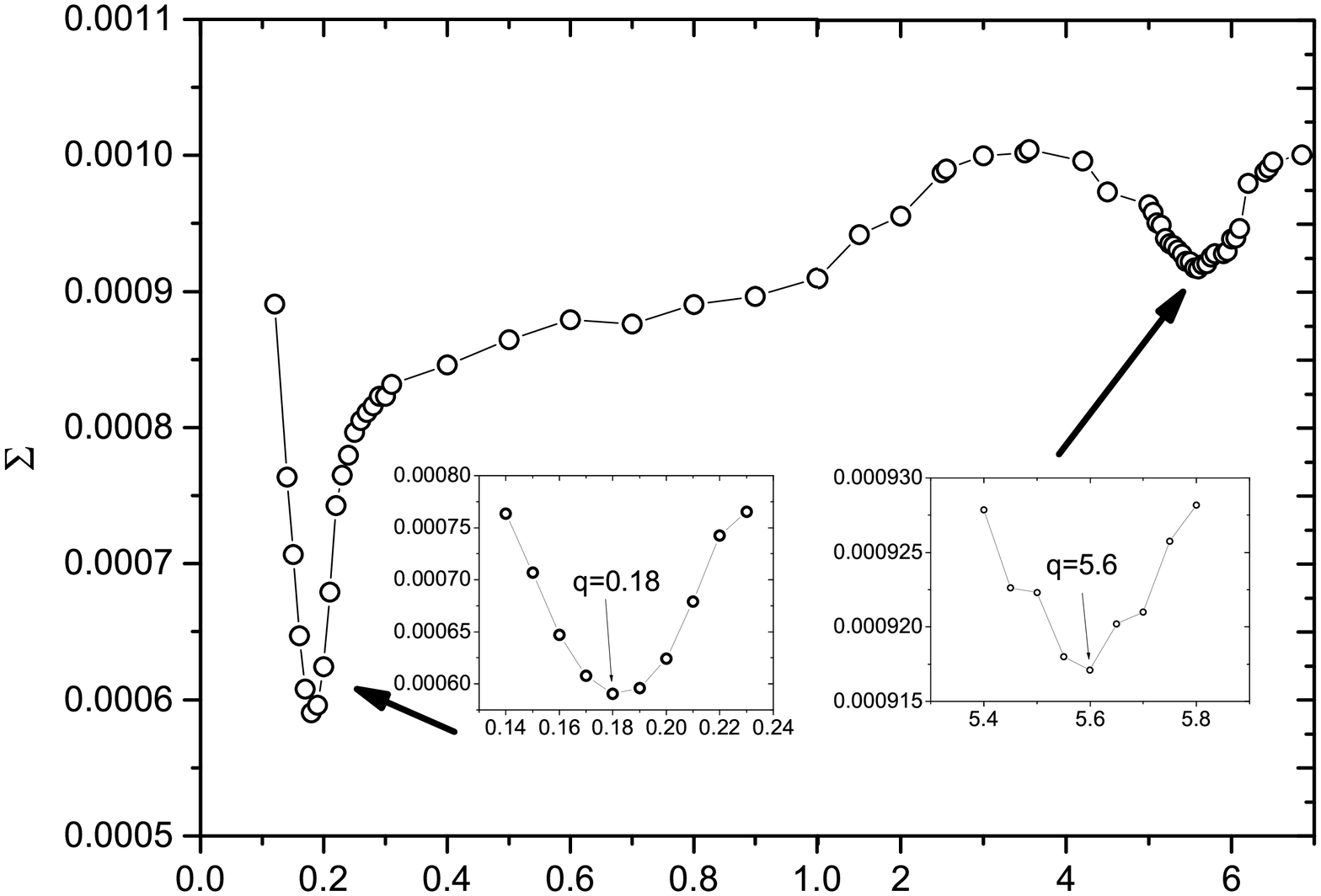}
\caption{The relation between $\Sigma$ and $q$ for V1853 Ori, and the minimum is found at $q=0.18$.}\label{qsearch}
\end{center}
\end{figure}

\begin{figure}
\begin{center}
\includegraphics[width=10cm]{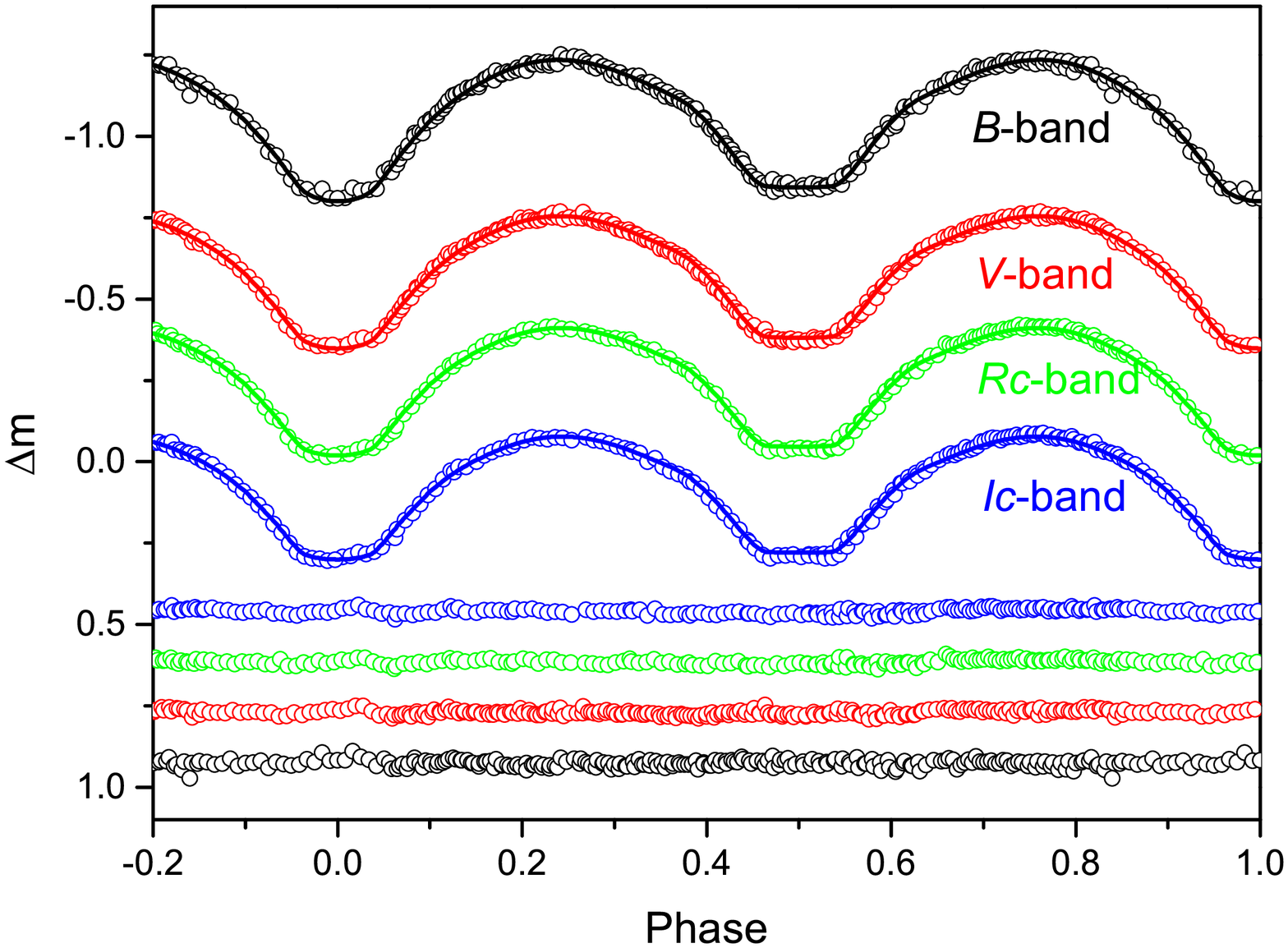}
\caption{Observed light curves of V1853 Ori in $B$, $V$ ,$Rc$ and $Ic$ bands and their fits by theoretical light curves by using the W-D code.}\label{oclc}
\end{center}
\end{figure}

\begin{figure}
\begin{center}
\includegraphics[width=10cm]{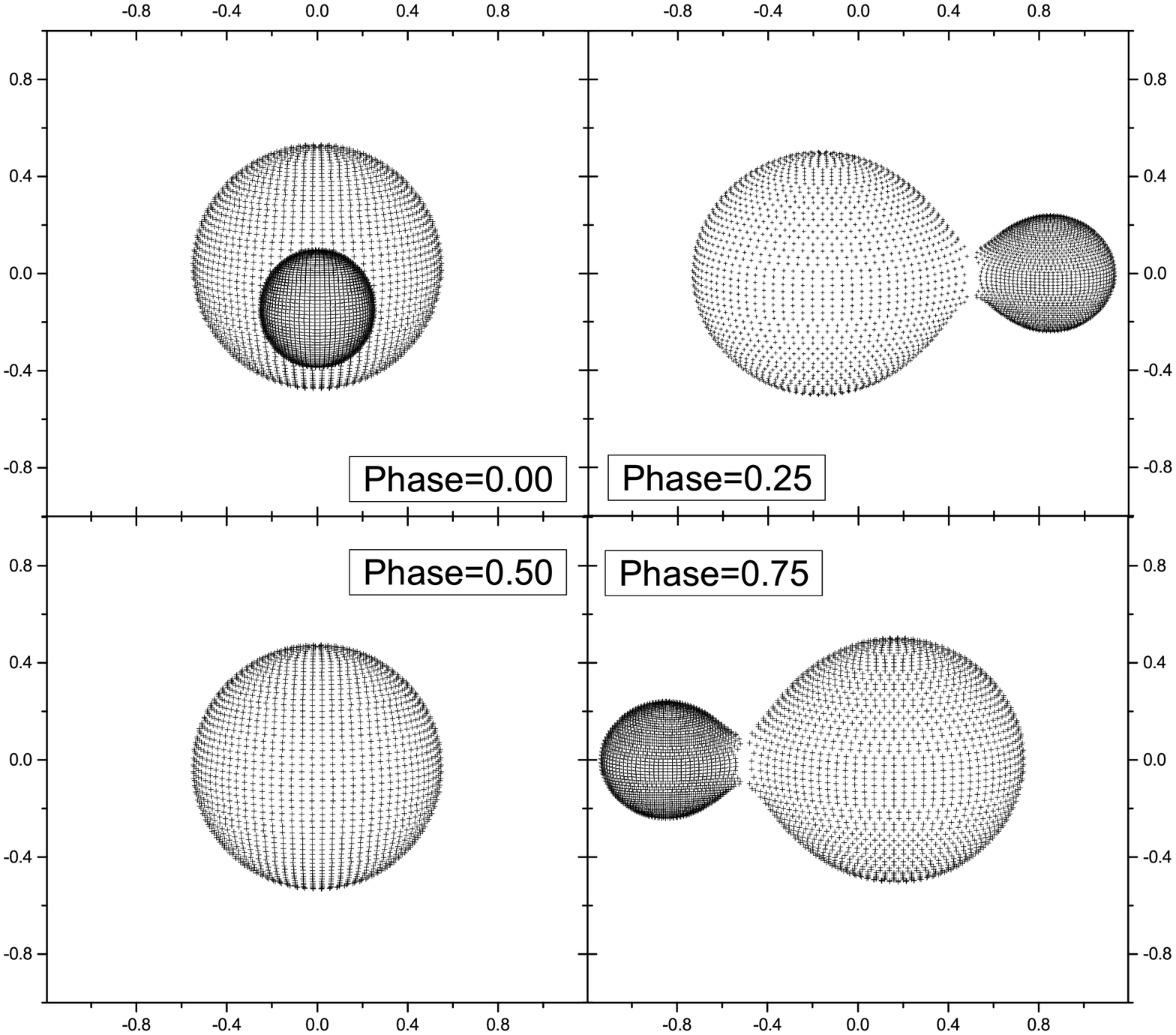}
\caption{ Geometrical structure of the A-type intermediate contact binary V1853 Ori at phases 0.00, 0.25, 0.50, and 0.75.}\label{structure}
\end{center}
\end{figure}

A $q$-search ($q=M_{2}/M_{1}$) method was used to determine the mass ratio of the system. Solutions were carried out for a series of values of the mass ratio. For each value of $q$, the calculation started at mode 2 (detached mode) and we found that the solutions usually converged to mode 3. The relation between the resulting sum $\Sigma$$W_{i}(O-C)_{i}^{2}$ of weighted square deviations and $q$ is plotted in Figure \ref{qsearch}. As shown in the figure, two minima are found at $q=0.18$ and $q=5.6$. They are the inverses of each other. The $\Sigma$$W_{i}(O-C)_{i}^{2}$ at $q=0.18$ is less than that at $q=5.6$. Therefore, we chose the initial value of $q$ as 0.18 and made it as an adjustable parameter. Then we performed a differential correction until it converged and final solutions were derived. The solution converged at $q=0.1896(0.0013)$. The photometric solutions are listed in column 2 Table \ref{phsolutions} and the theoretical light curves computed with those photometric parameters are plotted in Figure \ref{oclc}. The light curves are nearly symmetric and no spotted solution is needed. The solution reveals that V1853 Ori is A-type intermediate contact binary with a degree of contact factor of $f=33.3\%(3.7\%)$ and a mass ratio of $q=0.1896(0.0013)$. The geometrical structures at phases 0.0, 0.25, 0.5, and 0.75 are shown in Figure \ref{structure}.

\begin{table}
\caption{Photometric solutions analyzed using the W-D code.}\label{phsolutions}
\begin{center}
\small
\begin{tabular}{lcc}
\hline
Parameters                                                          & Photometric elements   \\\hline
orbital inclination i                                               &  80.19(24)             \\
mass ratio $m_{2}/m_{1}$                                            &  0.1896(13)            \\
primary temperature $T_{1}$                                         &  6240K                 \\
temperature ratio $T_{2}/T_{1}$                                     &  0.9995(11)            \\
Luminosity ratio $L_{1}/(L_{1} + L_{2})$ in $B$ band                &  0.81200(24)           \\
Luminosity ratio $L_{1}/(L_{1} + L_{2})$ in $V$ band                &  0.81214(22)           \\
Luminosity ratio $L_{1}/(L_{1} + L_{2})$ in $Rc$ band               &  0.81216(24)           \\
Luminosity ratio $L_{1}/(L_{1} + L_{2})$ in $Ic$ band               &  0.81219(27)           \\
Luminosity ratio $L_{2}/(L_{1} + L_{2})$ in $B$ band                &  0.18800(24)           \\
Luminosity ratio $L_{2}/(L_{1} + L_{2})$ in $V$ band                &  0.18786(22)           \\
Luminosity ratio $L_{2}/(L_{1} + L_{2})$ in $Rc$ band               &  0.18784(24)           \\
Luminosity ratio $L_{2}/(L_{1} + L_{2})$ in $Ic$ band               &  0.18781(27)           \\
Modified dimensionless surface potential of star 1                  &  2.1663(44)            \\
Modified dimensionless surface potential of star 2                  &  2.1663(44)            \\
fillout factor                                                      &  0.333(37)             \\
Radius of star 1 (relative to semimajor axis) in pole direction     &  0.50066(83)           \\
Radius of star 1 (relative to semimajor axis) in side direction     &  0.5490(12)            \\
Radius of star 1 (relative to semimajor axis) in back direction     &  0.5742(13)            \\
Radius of star 2 (relative to semimajor axis) in pole direction     &  0.2404(31)            \\
Radius of star 2 (relative to semimajor axis) in side direction     &  0.2516(38)            \\
Radius of star 2 (relative to semimajor axis) in back direction     &  0.2955(83)            \\
Equal-volume radius of star 1 (relative to semimajor axis) $R_{1}$  &  0.54201(63)           \\
Equal-volume radius of star 2 (relative to semimajor axis) $R_{2}$  &  0.2634(30)            \\
Radius ratio $R_{2}/R_{1}$                                          &  0.4859(56)            \\
$\Sigma{\omega(O-C)^2}$                                             &  0.000595              \\
\hline
\end{tabular}
\end{center}
\end{table}

\section{Discussions and Conclusions}

\begin{figure}
\begin{center}
\includegraphics[width=10cm]{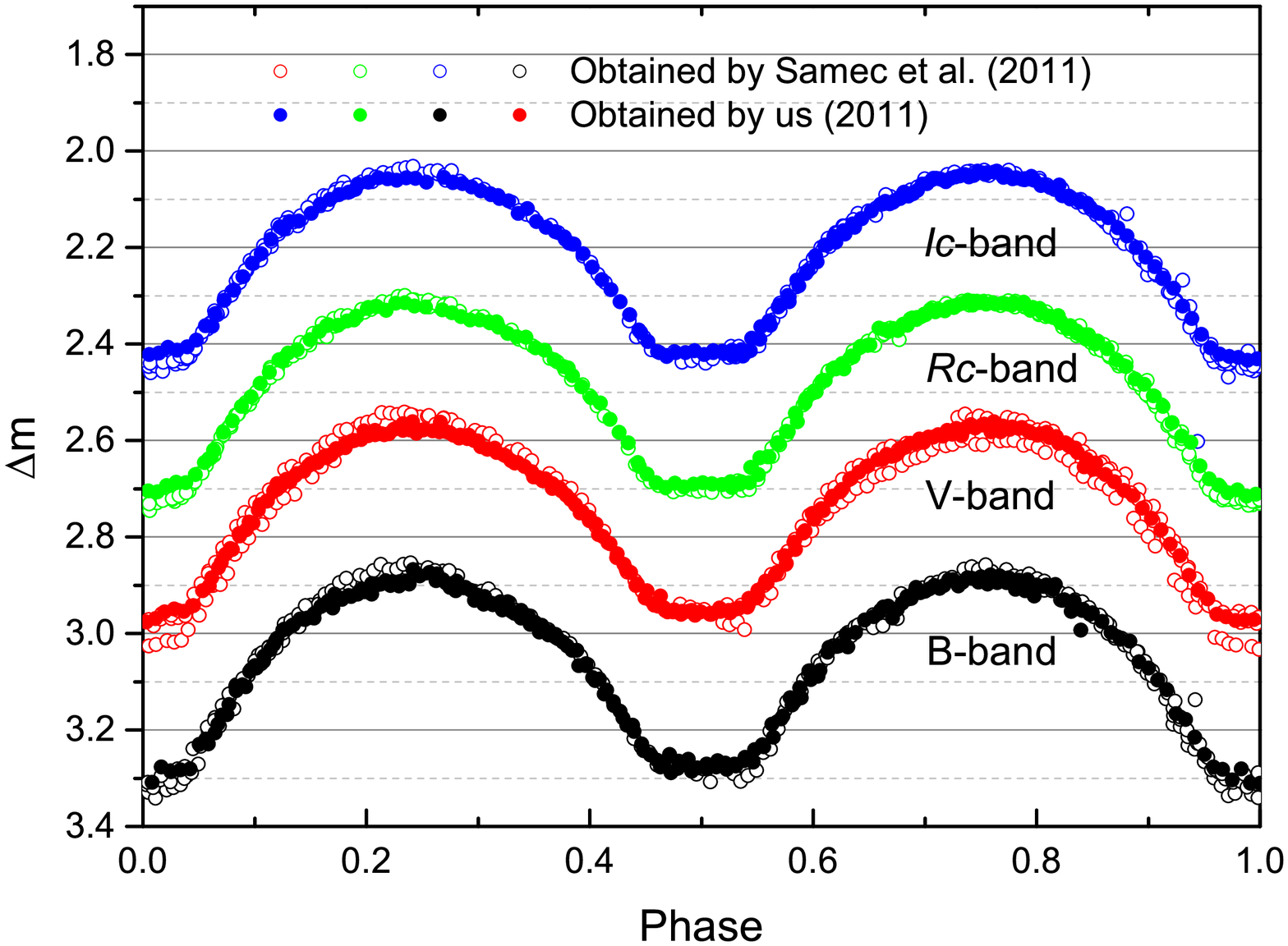}
\caption{The comparison of the two sets $B$, $V$, $Rc$ and $Ic$ light curves obtained by \cite{2011AJ....142..117S} and by us. The $open$ $circles$ refer to the data obtained by \cite{2011AJ....142..117S}, the $dots$ refer to the data obtained by us.}\label{compareLC}
\end{center}
\end{figure}

The light curves of V1853 Ori published by \cite{2011AJ....142..117S} are asymmetric and showing O'Connell effect. However, when we observed it in 2012, the light curves are symmetric and the O'Connell effect disappear. The comparison of the two sets of light curves is displayed in Fig. \ref{compareLC}. It is shown that the maxima at phase 0.25 in the light curves of \cite{2011AJ....142..117S} are slightly higher than the other ones at phase 0.75, while the two maxima in our light curves are nearly equal. Moreover, as shown in the figure, the light curves obtained by \cite{2011AJ....142..117S} displayed a larger scatter. These indicate that our light curves could be used to determine more reliable solutions.
\cite{2011AJ....142..117S} used two dark spots on the primary component to explain the asymmetries of their light curves. Their results shows that V1853 Ori is an extreme mass ratio ($q=0.2$), W-type contact binary with a fill-out factor of $\sim$50\%. By analyzing our light curves obtained by the 1.0-m reflecting telescope at the Yunnan Observatories with the W-D code, we found that V1853 Ori is an A-type contact binary with a mass ratio $q$=0.1896 that is slightly smaller than the value $q$=0.20 obtained by \cite{2011AJ....142..117S}. But the fill-out factor we derived is 33.3\% that is much smaller than that 50\% obtained by \cite{2011AJ....142..117S}. The photometric results of \cite{2011AJ....142..117S} showed that the temperature of the less-massive component was about 61\,K higher than the more-massive one, while our photometric results showed that the temperature of the less-massive is only 3\,K lower than the more-massive one. We think that our results are more credible because they are based on the hight-precision $B$, $V$, $Rc$ and $Ic$ bands and symmetric CCD LCs. It is possible that the two cool magnetic spots on the primary component are disappeared when we observed it. Maybe the asymmetric light curves have affected the photometry results.
The conclusions, the W-type, obtained by \cite{2011AJ....142..117S} may be caused by the two cool spots modeled on the primary component. We think that exchange between the A-type and the W-type can be explained by activity of some dark spot on the primary component, as pointed out by \cite{1964AJ.....69..154B}. Such phenomena also encountered in other W UMa binaries such as RZ Com (\citealt{2008ChJAA...8..465H}) and FG Hya (\citealt{2005MNRAS.356..765Q}).

The absolute magnitude of V1853 Ori in $V$ band is estimated as 3.67\,mag by using the relation $M_{V} = m_{V}-5log(1000/\pi)+5-A_{V}$ (e.g., \cite{2018ApJ...859..140C}), where the values of $m_{V}$, the parallax $\pi$ and $A_{V}$ are chosen as 11.60(0.16)\,mag (\citealt{2000A&A...355L..27H}), 2.6856(0.0465)\,mas (\citealt{2018A&A...616A...1G}) and 0.075(0.025)\,mag (\citealt{2018ApJ...859..140C}), respectively. The period and luminosity of V1853 Ori are in good agreement with the period-luminosity relation given by \cite{2018ApJ...859..140C}. The total luminosity of the two components should be 2.964 L$_{\odot}$ based on the equation $M_{bol} = -2.5logL/L_{\odot}+4.74$ and the bolometric correction $BC = M_{bol}-M_{V}$ = -0.11 (\citealt{2000asqu.book..381D}), where L$_{\odot}$ = 3.845$\times$$10^{33}$ ergs s$^{-1}$ and +4.74 is the absolute bolometric magnitude of the Sun. The luminosity of the primary component is calculated to be 2.964 $\times$ 0.81214 = 2.407 L$_{\odot}$, while that of the second component is 0.557 L$_{\odot}$. The mass of the primary component should be 1.20 M$_{\odot}$ based on the equation $logL/L_{\odot}$ = $3.8logM/M_{\odot}$+0.08 (\citealt{2000asqu.book..381D}). Then the mass of the second component is calculated as $M_2$ =0.23 M$_{\odot}$ based on our mass ratio. By using the kepler's third law, the semi-major axis of the binary is derived as 2.50 R$_{\odot}$. Finally, the radii of the components $R_{1}$ and $R_{2}$ are computed as 1.36 R$_{\odot}$ and 0.66 R$_{\odot}$, respectively.

By monitoring V1853 Ori for about six year, 10 times of light minimum were obtained. A period analysis with all available eclipse times shows that the period of V1853 Ori is decreasing at a rate of $dP/dt=-1.96(0.46)\times10^{-7}$ d yr$^{-1}$. This is very common in contact binary, such as V524 Mon ($dP/dt=-5.55\times10^{-8}$ d yr$^{-1}$, \citealt{2012PASJ...64...85H}), V532 Mon ($dP/dt=-1.716\times10^{-7}$ d yr$^{-1}$, \citealt{2016AJ....152..120H}), V1073 Cyg ($dP/dt=-2.24\times10^{-7}$ d yr$^{-1}$, \citealt{2018RAA....18...20T}). The long-term decrease of the orbital period could be explained by mass transfer from the more-massive component to the less-massive one. As the orbital period is decreasing, V1853 Ori will evolve into a high fill-out overcontact binary. Finally, it will merge and produce a luminous red nova similar to that of V1309 Sco (e.g., \citealt{2016RAA....16...68Z}).

\begin{acknowledgements}
This work is partly supported by the the National Natural Science Foundation of China (No. 11503077). CCD photometric observations of V1853 Ori were obtained with 0.4-m telescope at Naresuan University in Thailand, the 1-m and the 60-cm R-C reflect telescope, Yunnan Astronomical Observatory of Chinese Academy of Sciences.\\
\end{acknowledgements}

\label{lastpage}
\end{document}